# TEMPERATURE CONTROL STABILITY DURING SPARK PLASMA SINTERING

Charles Manière[a], Geuntak Lee[a,b], Eugene A. Olevsky[a,c]*

(a) Powder Technology Laboratory, San Diego State University, San Diego, USA
(b) Mechanical and Aerospace Engineering, University of California, San Diego, La Jolla, USA
(c) NanoEngineering, University of California, San Diego, La Jolla, USA

**Abstract**

The stability of the proportional–integral–derivative (PID) control of temperature in the spark plasma sintering (SPS) process is investigated. The PID regulations of this process are tested for different SPS tooling dimensions, physical parameters conditions, and areas of temperature control. It is shown that the PID regulation quality strongly depends on the heating time lag between the area of heat generation and the area of the temperature control. Tooling temperature rate maps are studied to reveal potential areas for highly efficient PID control. The convergence of the model and experiment indicates that even with non-optimal initial PID coefficients, it is possible to reduce the temperature regulation inaccuracy to less than 4 K by positioning the temperature control location in highly responsive areas revealed by the finite-element calculations of the temperature spatial distribution.

* Corresponding author: **CM**: Powder Technology Laboratory, San Diego State University, 5500 Campanile Drive, San Diego, CA 92182-1323,
Ph.: (619)-594-2420; Fax: (619)-594-3599, *E-mail address*: cmaniere@mail.sdsu.edu

The spark plasma sintering (SPS) process is an emerging technique [1-5] able to efficiently densify a wide variety of materials [6] while preserving a fine microstructure [7-8]. This process is similar to hot pressing using graphite tools and a pulsed electric current allowing to attain high temperatures (up to and above 2000 °C), heating rates (up to 1000 K/min) and high pressures (up to 150 MPa). These conditions enable the complete densification of refractory materials [9-10] and the substantial reduction of the sintering duration [11]; the outcomes of a few minutes in SPS compare to hours in conventional sintering. These process features permit the minimization of the grain growth and preservation of sub-micron scale microstructures leading to higher mechanical properties [5-8]. The SPS process in its traditional arrangements is a stable and reproducible technology [5-6]. However, for complex component shapes [12-13] and large size configurations [14], this technology needs to be specially adjusted [15-17] in order to ensure temperatures/densification homogeneity and a stable temperature regulation. The finite element method (FEM) is often used to assess the expected thermal gradients and densification in-homogeneities [18-19] and to find optimized process conditions. Numerous electro-thermal models highlight the prominent role of the electric and thermal contacts in the temperature generation and distribution during SPS [20-26].

In this work, the origins of the SPS regulation difficulties are studied and revealed by an FEM simulation, and the routes towards process optimization are determined and verified experimentally. The PID regulation is strongly influenced by the heating response [27-29] and the lag between the area of the main heat generation (more responsive) and the area of heating control. This lag can be potentially influenced by several factors: the tooling dimensions and the thermal contact resistance (TCR) which decreases the heat flux between the heat generation and control areas [26]. All these potential factors are varied and studied in the conducted simulation. Usually, PID correction of the temperature regulation is based on a tuning of the PID coefficients [30]. In this work, the responsiveness map is calculated by the FEM simulations and utilized to locate the areas of the optimal regulation efficiency.

Our starting point is a coupled electro-thermal finite element model [26] of the SPS regular setup (fig. 1). The model includes the Joule heating equations (detailed in ref [25]) and a set of temperature dependent electric and thermal contacts (ECR and TCR are electric and thermal contact resistances, respectively) previously determined for the same tool size and conditions [31]. The temperature dependent material properties are reported in Table. 1. The applied thermal cycle is a temperature ramp of 200 K/min up to 600°C and 8 min of dwell. A constant pressure of 50 MPa is applied during the entire cycle. The verification experiment has been performed using the SPS system *Dr. Sinter SPSS-515 (SPS Syntex Inc., Japan)*. In the both experiment and simulation, a fully dense silver pellet is considered a processed sample.

The configuration (fig. 1) uses a high pressure enabling thick die (inner diameter 10 mm, external diameter 30mm) and a conductive sample. In these conditions, the heat generation is mainly located in the punches and the die heating occurs with a time lag (fig. 2a.) In this case the PID regulation shows a lot of oscillations of the electric current and temperature (fig. 3a.) The oscillations imply two types of temperature distribution behavior. When the electric current increases, for instance at 180 s (fig. 2a), the maximum temperature is located in the middle of the punches and the die heating takes place with a time lag. During the electric current decreasing stage (220 s, fig. 2a), the heat is mainly evacuated by thermal conduction in the punches in contact with the colder spacer. Under these conditions the die cooling has a time lag. This lag of the die heating and cooling is explained by the presence of a high TCR at the vertical punch/die contact [26, 31]. As shown in our previous works [24, 26, 31], the punch/die TCR is very high (about 1E-3 K.m$^2$/W) due to a low contact pressure and the presence of the graphite foil that possesses a low thermal conductivity along its thickness. During the heating stage, the TCR decreases the heat exchanges from the punches to the die. During the cooling stage, the heat flux that passes across this interface is reduced by the TCR, inducing the time lag in the cooling at the die surface.

This impact of the vertical interface TCR on the temperature regulation can be easily verified. If the TCR is multiplied by two, the punch/die temperature difference increases from 100 K (fig. 2a) to 200 K (fig. 2b) and both the temperature and electric current oscillations' magnitude increase (fig. 3a). In a similar way, if the die thickness is increased, the mass of the die material to be heated by the punches is increased and the punch/die temperature difference is also about 200 K (fig. 2c) causing high regulation oscillations (fig. 3b). We also tried to reduce the heat generation by punches by reducing their heights in order to homogenize the temperature and stabilize the PID regulation. The results reported in fig. 2d show that the punch/die temperature difference is reduced down to 80 K but the PID regulation still includes a lot of oscillations (fig. 3b).

As a rule, these PID regulation difficulties can be solved by tuning the PID coefficients or by placing the temperature control in a highly responsive area. The tuning of the PID coefficients generally requires long trial and error procedures. In this work, the finite element simulation is used to find the area within the SPS setup having the highest heating responsiveness. In order to reveal such areas, finding the heating rate field generated by the electric current impulse is a required key issue. Indeed, the higher the heating response is, the lower the heating time lag is, which improves the PID regulation. The heating rate field evolution for the die regulation is shown in fig. 4. In the beginning, in the middle and at the end of the heating ramp, the heating rate is maximal in the punches being about 10 K/s. In the beginning of the dwell time during the temperature oscillations, the punches still experience the maximum heating rate, but its values decrease to about 6 K/s. At the end of the dwell time after the stabilization of the PID regulation, the heating rate becomes close to zero. Therefore, the areas of maximum responsiveness are located not in the die but at the mid-height of the punches. As a result, the punches are a better candidate for the PID regulation compared to the die whose heating occurs with the time lag (see fig. 2a). Similarly, to the approach used by other authors [5, 16, 19 21, 28] with FCT SPS system [32], the PID regulation is now located in the upper punch, at its mid-height (see fig. 1). The regulation temperatures curves are reported in fig. 4. The regulation which provided far from optimal results when associated with the die (red curve, fig. 4), is now nearly perfect when the temperature is measured in the punch (black and blue dashed curves, fig. 4). The presence of the temperature overshot at the beginning of the dwell time is almost removed. The simulated overshot is 3.1 K and the experimentally measured overshot is 4.2 K.

In conclusion, the instabilities of the SPS PID regulation are originated from the heating time lag between the electric current impulsion and the heat response. It is shown that this lag is minimized in the punches where the majority of the heat is generated and confined, due to the presence of a high punch/die TCR. In contrast, the die heating and cooling experience time lag which generates high PID regulation difficulties. When the PID control is placed in the area of the maximum responsiveness revealed by the FEM calculation, the PID regulation follows the prescribed temperature cycle more accurately. The overshot at the beginning of the dwell time, which was reported in multiple SPS studies, is thereby removed. The temperature error caused by PID control is reduced down to 4 K. This case study shows that finding the optimal location of the temperature control using FEM temperature distribution simulations is a more efficient way to optimize the temperature regulation compared to tuning the PID coefficients requiring time-consuming trial and error procedures.

**Acknowledgements**

The support of the US Department of Energy, Materials Sciences Division, under Award No. DE-SC0008581 is gratefully acknowledged.

## Figure captions

Fig. 1: Cross-section of the SPS showing the location of each electric/thermal horizontal/vertical contact resistance (CR) taken into account in the model, the main thermal boundary conditions, and the two different temperature control locations.

Fig. 2: Temperature fields determined for the die controlled regulation: a) at 180 and 220s, b) with vertical thermal contact resistance (TCR) multiplied by 2, c) with thicker die, d) with smaller punches.

Fig. 3: PID regulated temperature (upper) and electric current (lower) under different conditions: a) die temperature control regulation with vertical TCR multiplied by 2, b) die temperature control regulation for small punch and thick die configurations.

Fig. 4: Die/punch PID regulation and responsiveness map during the SPS process cycle.

------------------

Graphical abstract: Overview of the optimized PID control study for the spark plasma sintering (SPS) technology, with on the left, the Joule heating simulation temperature field at the beginning of the dwell, in the middle, the regulated temperature curves in die/punch control modes and on the right, the heating rate field that help revealing the optimal location for the PID control measurement ; it is shown the simple identification of the high thermal responsive area of the SPS column enable a drastic improvement of the PID regulation quality (regulation overshot <4 K) without any PID coefficient tuning.

**Table captions**

Table 1: Electro-thermal materials parameters used in FEM calculations.

# Table

Table 1: Electro-thermal materials parameters used in FEM calculations.

| Materials | | Expression |
|---|---|---|
| Graphite | $Cp$ ($J.kg^{-1}.K^{-1}$) | $34.3+2.72.T-9.6E-4.T^2$ |
| Electrode | | $446.5+0.162.T$ |
| Silver | | $226+0.0171.T+5.01E-5.T^2-1.77E-8.T^3$ |
| Graphite | $\kappa$ ($W.m^{-1}.K^{-1}$) | $123-6.99E-2.T+1.55E-5.T^2$ |
| Electrode | | $9.99+0.0175.T$ |
| Silver | | $420+0.0998.T-2.94E-4.T^2+2.11E-7.T^3-5.79E-11.T^4$ |
| Graphite | $\rho$ ($kg.m^{-3}$) | $1904-0.0141.T$ |
| Electrode | | $7900$ |
| Silver | | $10659-0.469.T-2.98E-4.T^2+1.47E-7.T^3$ |
| Graphite | $\rho_e$ ($\Omega.m$) | $1.70E-5-1.87E-8.T+1.26E-11.T^2-2.46E-15.T^3$ |
| Electrode | | $(50.2+0.0838.T-1.76E-5.T^2).1E-8$ |
| Silver | | $8.27E-18.T^3-3.08E-15.T^2+6.07E-11.T-1.81E-09$ |

# Figures

Fig. 1: Cross-section of the SPS showing the location of each electric/thermal horizontal/vertical contact resistance (CR) taken into account in the model, the main thermal boundary conditions, and the two different temperature control locations.

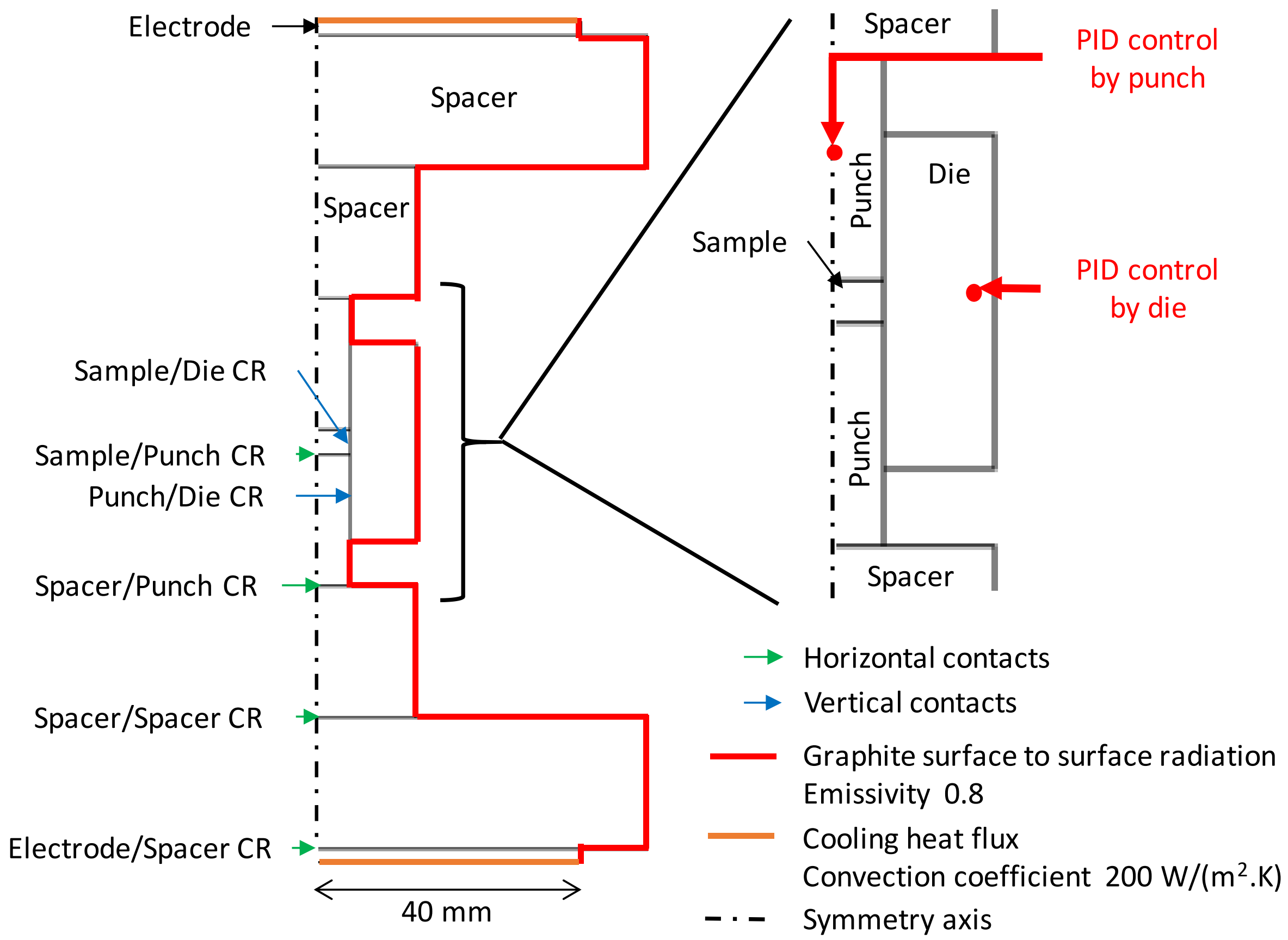

Fig. 2: Temperature fields determined for the die controlled regulation: a) at 180 and 220s, b) with vertical thermal contact resistance (TCR) multiplied by 2, c) with thicker die, d) with smaller punches.

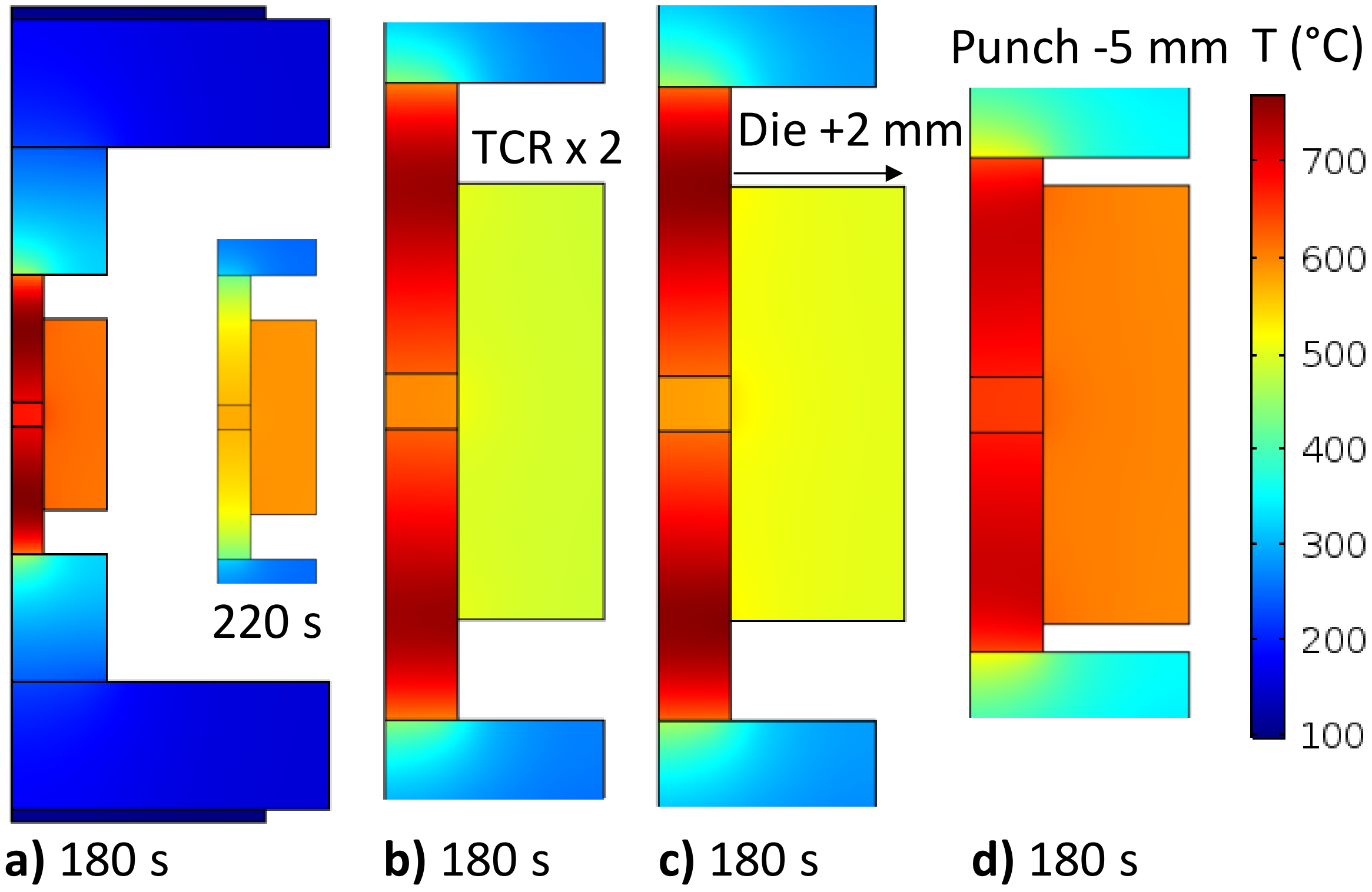

Fig. 3: PID regulated temperature (upper) and electric current (lower) under different conditions: a) die temperature control regulation with vertical TCR multiplied by 2, b) die temperature control regulation for small punch and thick die configurations.

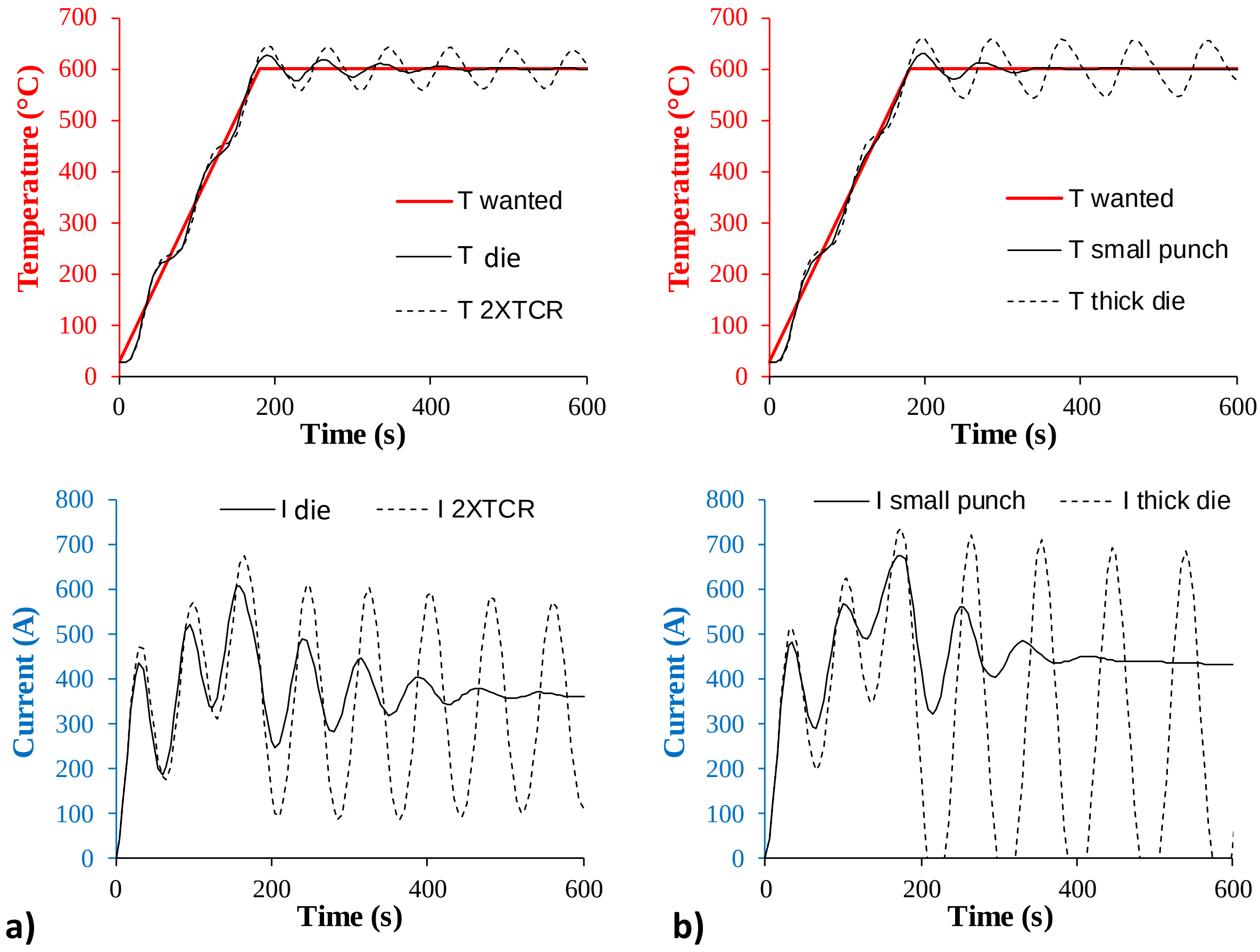

Fig. 4: Die/punch PID regulation and responsiveness map during the SPS process cycle.

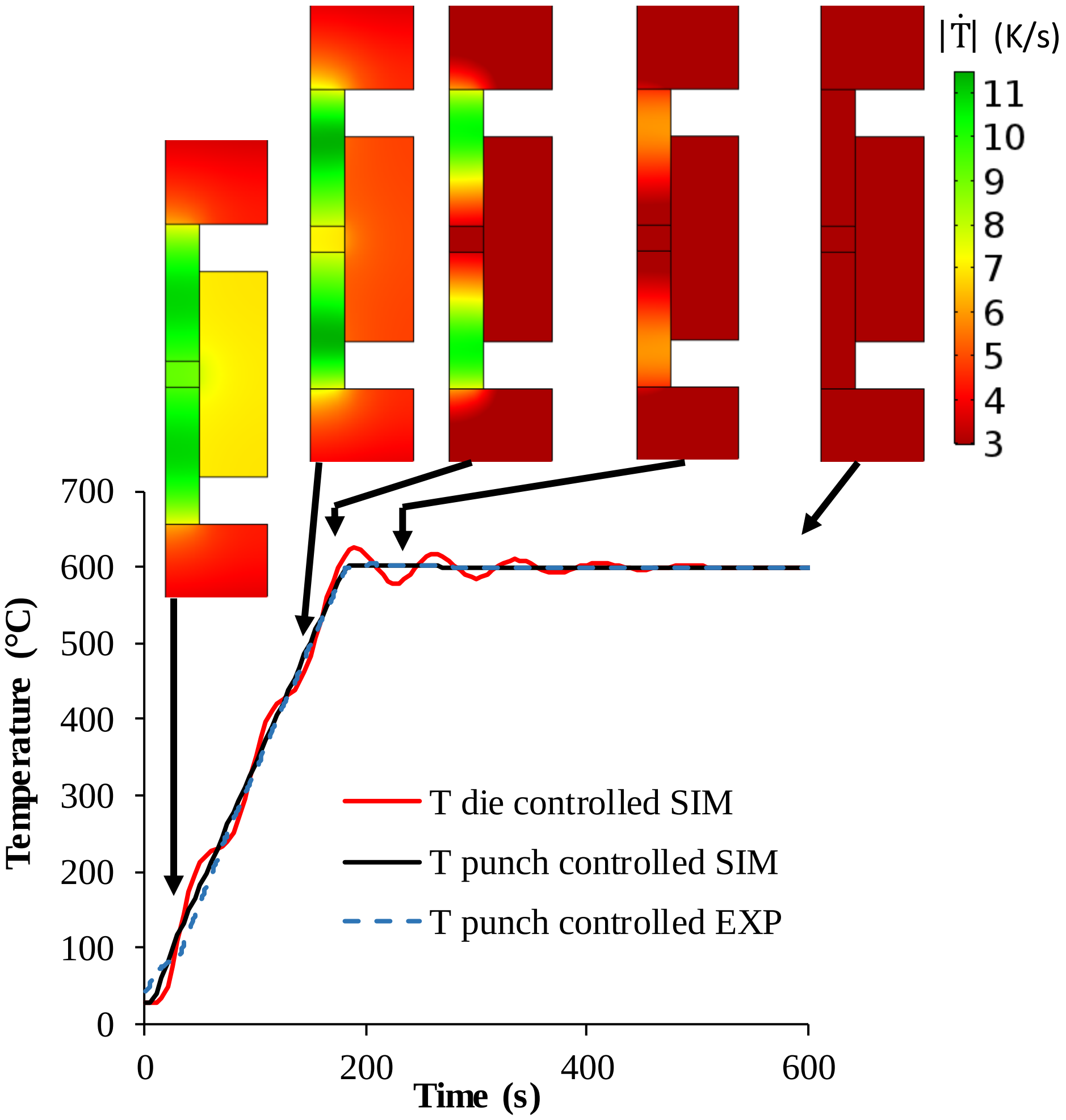

# Graphical Abstract

Graphical abstract: Overview of the optimized PID control study for the spark plasma sintering (SPS) technology, with on the left, the Joule heating simulation temperature field at the beginning of the dwell, in the middle, the regulated temperature curves in die/punch control modes and on the right, the heating rate field that help revealing the optimal location for the PID control measurement ; it is shown the simple identification of the high thermal responsive area of the SPS column enable a drastic improvement of the PID regulation quality (regulation overshot <4 K) without any PID coefficient tuning.

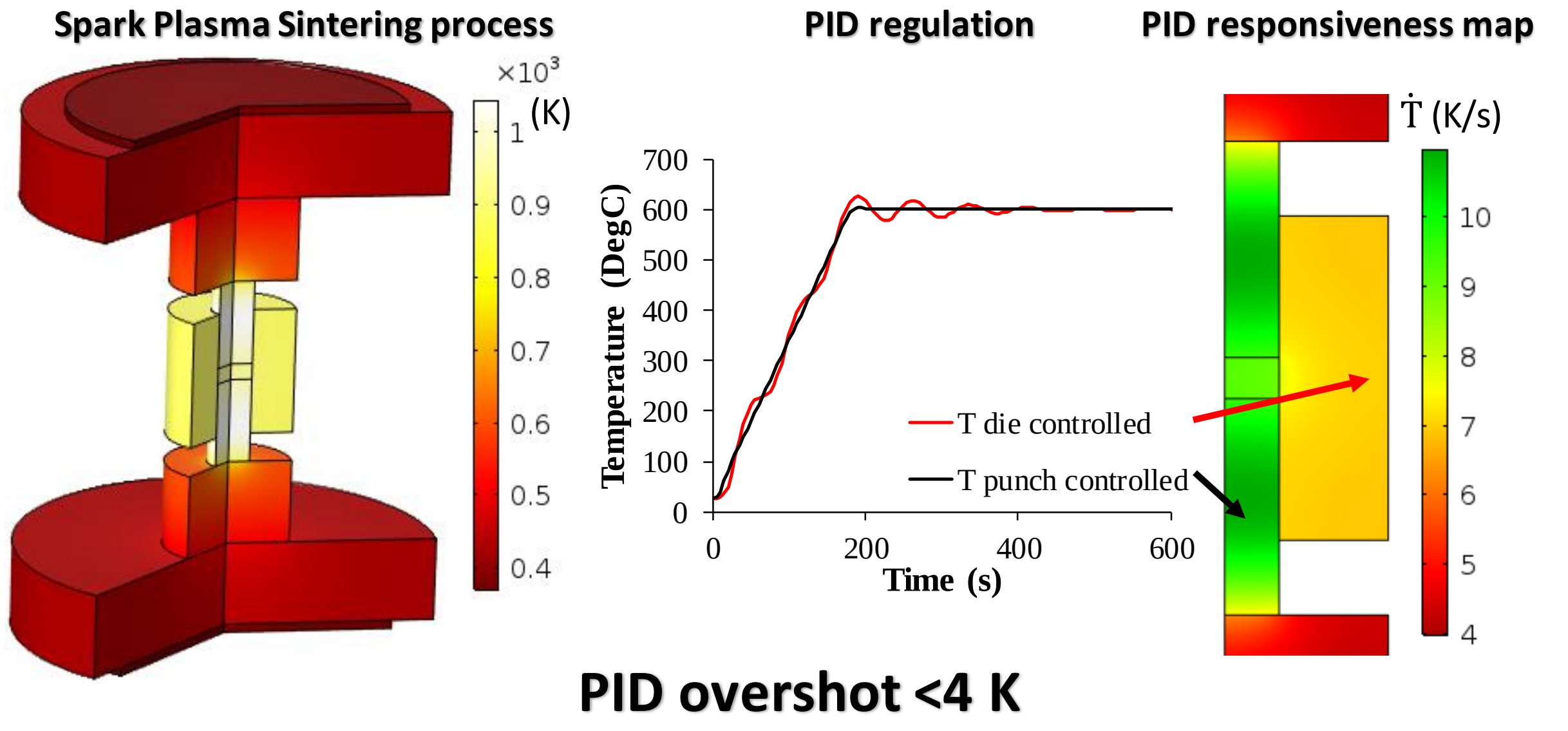